\documentclass{osa-article}

\journal{oe}

\articletype{Research Article}
\usepackage{longtable}
\usepackage{appendix}
\usepackage{titlesec}
\usepackage{capt-of}
\usepackage{arydshln}
\usepackage{todonotes}
\usepackage[english]{babel}
\usepackage{blindtext}
\usepackage{pstricks}
\usepackage{breqn}
\usepackage{comment,psfrag,subfigure,enumerate}
\usepackage{cancel}
\usepackage{cite}
\usepackage{url}
\usepackage{tensor}
\usepackage{graphicx}
\usepackage{pstool}
\usepackage{float,color}
\usepackage{rotating}
\usepackage{multirow}
\usepackage{multicol}
\usepackage{setspace}
\usepackage{hyperref}
\usepackage{graphicx} 
\usepackage{epstopdf} 
\usepackage{subfigure}

\usepackage{soul}

\usepackage{times}
\usepackage{tikz}
\usepackage{verbatim}
\usepackage{pst-optexp}
\usepackage[framemethod=TikZ]{mdframed}

\usepackage{navigator}
\usepackage{pgfplotstable,enumerate}
\usetikzlibrary[pgfplots.groupplots]
\usepgflibrary{decorations.pathmorphing} 
\usetikzlibrary{spy}
\usetikzlibrary{arrows.meta}
\usetikzlibrary{arrows,shapes}
\usetikzlibrary{arrows,patterns}

\newcommand{\Rmnum}[1]{\expandafter\@slowromancap\romannumeral #1@}

\usetikzlibrary{calc}
\mdfdefinestyle{MyFrame}{%
    usetwoside=false,
    linecolor=black,
    outerlinewidth=0.5pt,
    roundcorner=10pt,
    innertopmargin=\baselineskip,
    innerbottommargin=\baselineskip,
    innerrightmargin=10pt,
    leftmargin=10pt,
    rightmargin=10pt,
    skipabove = -10pt,
    skipbelow = 0pt,
    innerleftmargin=10pt,
    userdefinedwidth = 1.0\textwidth
    backgroundcolor=gray!1!white}
    
    \mdfdefinestyle{MyFrame1}{%
    usetwoside=false,
    linecolor=black,
    outerlinewidth=0.5pt,
    roundcorner=10pt,
    innertopmargin=\baselineskip,
    innerbottommargin=\baselineskip,
    innerrightmargin=10pt,
    leftmargin=10pt,
    rightmargin=10pt,
    skipabove = 10pt,
    skipbelow = 0pt,
    innerleftmargin=10pt,
    userdefinedwidth = 1.0\columnwidth
    backgroundcolor=gray!1!white}

\newcolumntype{f}{>{$}l<{$}}
\newcolumntype{n}{l}
\newcolumntype{N}{>{\scriptsize}l}
\newcolumntype{v}[1]{>{\raggedright\hspace{0pt}}p{#1}}
\newcolumntype{V}[1]{>{\scriptsize\raggedright\hspace{0pt}}p{#1}}
\newcolumntype{B}[1]{>{\boldmath\DC@{.}{,}{#1}}l<{\DC@end}}
\newcolumntype{d}[1]{>{\DC@{.}{,}{#1}}l<{\DC@end}}
\newcolumntype{i}[1]{>{\DC@{.}{,}{#1}\mathnormal\bgroup}l<{\egroup\DC@end}}
\newcolumntype{s}[1]{>{\DC@{.}{,}{#1}\mathsf\bgroup}l<{\egroup\DC@end}}
\newcolumntype{R}[1]{%
  >{\begin{turn}{90}\begin{minipage}{#1}\scriptsize\raggedright\hspace{0pt}}l%
  <{\end{minipage}\end{turn}}%
}
\newcolumntype{x}{>{\scriptsize\raggedright\hspace{0pt}}X}
\makeatother

\usetikzlibrary{intersections,arrows,decorations.pathmorphing,snakes}

\pgfdeclaredecoration{fiber}{initial}
{
    \state{initial}[width=1.5\pgfdecorationsegmentamplitude, next state=final]{
    \pgfpathmoveto{\pgfpointorigin}
    \pgflineto{\pgfpoint{\pgfdecoratedinputsegmentlength/2}{0pt}}
    \pgfpathcircle{\pgfpoint{\pgfdecoratedinputsegmentlength/2}{.5\pgfdecorationsegmentamplitude}}{.5\pgfdecorationsegmentamplitude}
    \pgfpathcircle{\pgfpoint{\pgfdecoratedinputsegmentlength/2-.25\pgfdecorationsegmentamplitude}{.5\pgfdecorationsegmentamplitude}}{.5\pgfdecorationsegmentamplitude}
    \pgfpathcircle{\pgfpoint{\pgfdecoratedinputsegmentlength/2+.25\pgfdecorationsegmentamplitude}{.5\pgfdecorationsegmentamplitude}}{.5\pgfdecorationsegmentamplitude}
    \pgfpathmoveto{\pgfpoint{\pgfdecoratedinputsegmentlength/2}{0pt}}
    }
   \state{final}{
        \pgfpathlineto{\pgfpointdecoratedpathlast}
   }
}

\begin{document}

\title{Optical Channel Impulse Response-Based Localization Using An Artificial Neural Network}
\author{Hamid~Hosseinianfar\authormark{1,*}~\jumplink{https://orcid.org/0000-0001-7998-138X}{\includegraphics[scale=0.05]{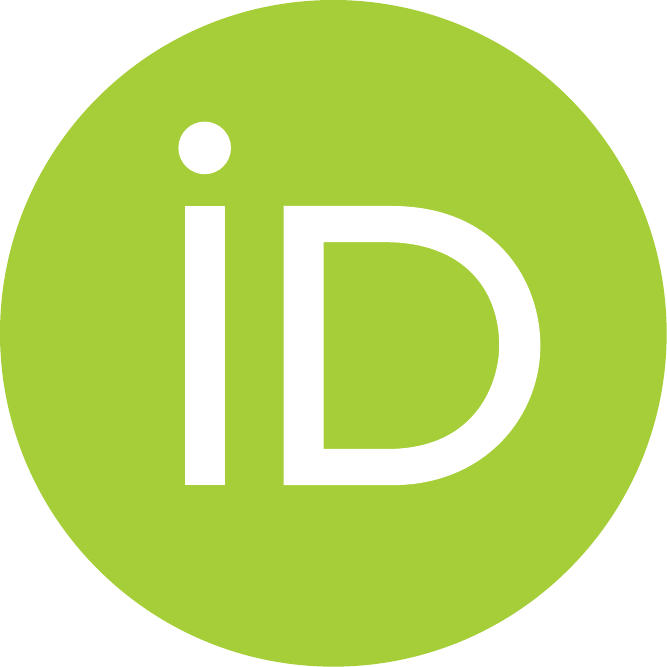}}, Hami~Rabbani\authormark{2}~\jumplink{https://orcid.org/0000-0003-3532-469X}{\includegraphics[scale=0.05]{Figures/ORCID-iD.pdf}}, Ma\text{\"i}t\'{e}~Brandt-Pearce\authormark{3}~\jumplink{https://orcid.org/0000-0002-2566-8280}{\includegraphics[scale=0.05]{Figures/ORCID-iD.pdf}}}

\address{\authormark{1,2,3} Charles L. Brown Department of Electrical and Computer Engineering, University of Virginia, Charlottesville, VA 22904 (e-mails: hh9af, dmr3ub, mb-p@virginia.edu). The corresponding author is Hamid Hosseinianfar (e-mail: hh9af@virginia.edu).}









\begin{abstract}
Visible light positioning has the potential to yield sub-centimeter accuracy in indoor environments, yet conventional received signal strength (RSS)-based localization algorithms cannot achieve this because their performance degrades from optical multipath reflection. However, this part of the optical received signal is deterministic due to the often static and predictable nature of the optical wireless channel. In this paper, the performance of optical channel impulse response (OCIR)-based localization is studied using an artificial neural network (ANN) to map embedded features of the OCIR to the user equipment's location. Numerical results show that OCIR-based localization outperforms conventional RSS techniques by two orders of magnitude using only two photodetectors as anchor points. The ANN technique can take advantage of multipath features in a wide range of scenarios, from using only the DC value to relying on high-resolution time sampling that can result in sub-centimeter accuracy.
\end{abstract}



\section{Introduction}
\todo[inline,caption={},disable]{List of tasks remained:\begin{itemize}
    \item Fig3.: generating the Tikz fig.
    \item Introduction: revising for the sake of coherency (redundant parts can be moved to related work Section)
    \item Description of ANN and hyper-parameter tuning
    \item Numerical results: red challenges and future work
\end{itemize}}
While the global positioning system (GPS) offers a ubiquitous localization service for outdoor environments, it is not a perfect match for indoor applications due to signaling coverage and reflections, resulting in low availability and accuracy for indoor spaces. Other RF techniques, like Wi-Fi-based localization, intrinsically lack localization accuracy proportional to their wavelength. Visible light communications (VLC)-based techniques are being developed that can provide indoor localization with greater precision than Wi-Fi and other RF technologies. VLC localization has also attracted substantial attention due to the broad use of light-emitting diodes (LED) for  indoor lighting. Optical channel impulse response (OCIR)-based  VLC positioning methods benefit from the largely static and deterministic optical wireless channel for localization \cite{brandt2020positioning,Performance_limit_hosseinianfar,7996815,8647875,8959201}. This paper uses an artificial neural network (ANN) to evaluate the effectiveness of employing  multipath reflection information within the OCIR with varying levels of temporal resolution for localization.

Even with significant signal processing, RF-based approaches have not been able to localize with under-decimeter level accuracy. Most research studies on RF-localization/sensing focus on developing experimental algorithms on off-the-shelf Wi-Fi network cards using the received signal strength (RSS) indicator \cite{depatla2015occupancy,8480076} or channel state information \cite{konings2018device,vasisht2016decimeter,ayyalasomayajula2020deep}. These researchers came to the conclusion that experimental investigations are fundamental to solving sensing/localization problems in realistic scenarios. Small improvements in accuracy also necessitate an exponential increase in both hardware (number of access-points and multiple input multiple output antennas) and
computational complexity. 

\todo[inline,caption={},disable]{\begin{itemize}
    \item CSI-based approaches \cite{shi2018accurate}: The RF CSI is unpredictable. This paper introduces a probabilistic fingerprinting technique as an accurate solution. Their scheme further boosts the localization efficiency by using principal component analysis to filter the most relevant feature vectors. Furthermore, with Bayesian filtering, we continuously track the trajectory of the moving subject. Accuracy Claim: need more investigation (vague results)
    \item \cite{DarConFerGioWin:J09,MarGifWymWin:J10}: UWB localization
\end{itemize}}
\todo[inline,disable]{Prof. Comment: Conclude the paragraph with "none of these approaches is able to locate an object within less than xx cm"}

{Optical wireless-based sensing and localization demonstrate superior performance based on theoretical limits \cite{wang2013position,zhang2014theoretical}, heuristic \cite{al2019lidal} and experimental \cite{6950776,li2016practical} methods. However, these techniques do not take advantage of multipath reflection patterns to reach the ultimate localization accuracy.}
In our initial research studies \cite{7996815, Performance_limit_hosseinianfar}, we  presented a proof of concept and fundamental limits for OCIR-based localization using a simple nearest neighbor algorithm for localization. In \cite{7996815}, we proposed an infrared uplink positioning algorithm to identify the user location information from the multipath reflections as captured by the fingerprinting map of a small selection of features of the OCIR.  However, the localization potential using high temporal resolution of the OCIR is an open problem.

In this paper, we investigate the accuracy of OCIR-based localization by exploiting ANN for comprehensive feature extraction and prediction. We consider samples of the entire line-of-sight (LOS) and non-LOS (NLOS) impulse response as input to the ANN and estimate the location of the user in a two-dimensional space. The limits on OCIR temporal resolution due to practical limitations, such as the number of photodetectors, optical pulse-width and receiver sampling rate are taken into account. 

We benchmark the OCIR-based performance against the conventional linear trilateration \cite{gu2016impact}.  We show that our algorithm benefits from multipath reflection even when using limited resolution hardware, i.e., a receiver sampling rate of $50~\si{Msps}$ and transmitter pulse-width of $10~\si{ns}$. 
The high temporal resolution OCIR-localization performance is also compared with a simple ANN-enabled algorithm that uses the DC RSS from three photodetectors. 
Numerical results show that conventional trilateration is at a disadvantage compared with OCIR-based localization. The OCIR-based localization outperforms the conventional method by over one order of magnitude, even with a smaller number of anchor points. Numerical results further show that the OCIR algorithm with only one photodetector can outperform an ANN-enabled technique using three DC RSS measurements. 
\todo[inline,disable,caption={}]{\begin{itemize}
    \item We compare the OCIR-based algorithm with conventional received signal strength (RSS)-base.
\end{itemize}}

The rest of this paper is organized as follows. We contextualize our work with respect to other related research in Section~\ref{related_work}. In Section \ref{system.model}, we present the system model, consisting of the indoor localization system together with an optical wireless channel. Section~\ref{DL.Sec} describes artificial neural networks used for regression problems. Numerical results and discussion are provided in Section~\ref{Numerical.Sec}. Section~\ref{Conclusion.Sec} concludes the paper.

\section{Related Work}\label{related_work}




Active localization techniques often employ the RSS as their measurement, making these systems the most affordable choice.
Trilateration-based methods rely on finding the distance from the  user equipment (UE) to three anchor points based on a closed-form distance relation and a feature of the received signal, such as the RSS (angle of arrival \cite{6823667,sahin2015accuracy} or time difference of arrival \cite{6131130,6687128} can also be used). However, these methods have poor blocking performance and are sensitive to multipath reflection \cite{zhang2014theoretical, zhang2013comparison,csahin2015hybrid}. Therefore, the above-mentioned closed-form relation between the distance and the received signal feature is not always dependable. Fingerprinting-based methods, on the other hand, resolve this issue by creating a map of one or more specific features of the received signal and localizing the UE by capturing the received signal and searching for the closest match on this map \cite{lin2005performance,6244790}.

Visible light-based passive sensing techniques are categorized into two groups (adapted from \cite{wang2017passive}): (i) general visible light-based (unmodified lighting), assuming either a fully passive device-free object  \cite{yang2017ceilingsee,faulkner2019smart,8851195,faulkner2019watchers} or a device-equipped object \cite{zhang2016litell,8585067,hu2018lightitude,zhang2017pulsar}, and (ii) VLC-based techniques with an active transmitter and a device-free object \cite{li2015human,li2016practical,lascio2016localight,jarchlo2018li,hu2018efficient,8647875,nguyen2018eyelight,al2019lidal,hosseinianfar2020cooperative,majeed2021passive,MajeedIoT}.
Most of these sensing techniques focus on developing algorithms for high-level applications, such as occupancy detection \cite{jarchlo2018li,hosseinianfar2020cooperative,MajeedIoT}, gesture recognition \cite{li2015human,li2016practical}, and fall detection \cite{majeed2021passive}. A fully passive occupancy detection study with an accuracy beyond 90\% was conducted in
\cite{yang2017ceilingsee}, exploiting reverse-biased LED luminaires as photodetectors for sensing. Using an array of sensors on the wall, the same approach was adopted in \cite{faulkner2019smart}, capable of localizing an object by capturing the deviation in the RSS of the ambient light. An extended, efficient version of \cite{yang2017ceilingsee} was presented in \cite{8851195,faulkner2019watchers} resulting in decimeter-level localization accuracy. Since the RSS is not unique to a specific place, a high level of accuracy cannot be achieved via these techniques, as in their RF-based peers. However, VLC techniques tend to be superior to counterpart RF techniques due to the more predictable and static nature of light propagation indoors.

The OCIR has only been considered for indoor VLC-based localization in a few works. A passive visible light-based fall detection system was presented in \cite{majeed2021passive} to identify the state of a person, including upright or prone positions, in an indoor environment; a neural network is then employed to learn the relationship between the OCIR measurements and the state of the target.  \cite{MajeedIoT} showed that a passive indoor visible light positioning system is able to locate the object of interest by employing an ANN to process the OCIR. To the best of our knowledge, no study has addressed  active localization using the OCIR, where an ANN can estimate the UE position with sub-centimeter accuracy, as presented in this paper.

\section{System Description}\label{system.model}
\label{System_Description}
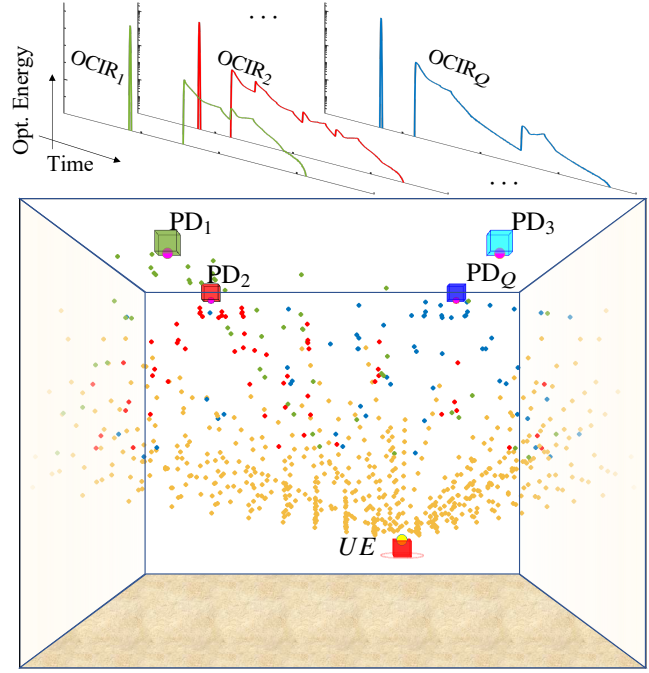
\begin{figure}[ht]
\centering
    \begin{tikzpicture}
		\node(image) at (0,0) {\includegraphics[width=3.4in]{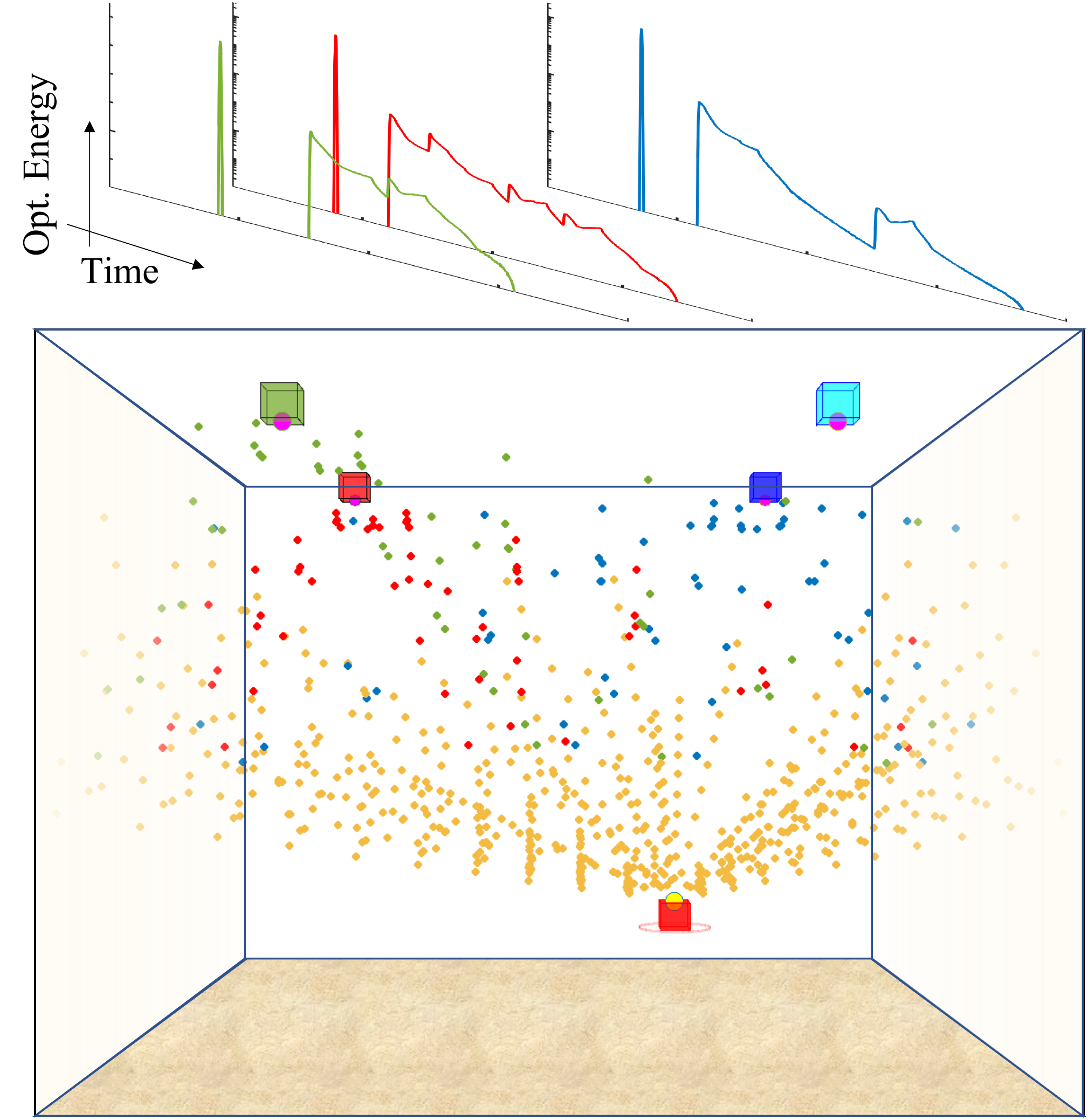}};
        \node[text width=3cm] at (3.4,0.8) {$\si{PD}_{Q}$};
        \node[text width=3cm] at (4,1.5) {$\si{PD}_{3}$};
        \node[text width=3cm] at (-0.05,0.8) {$\si{PD}_{2}$};
        \node[text width=3cm] at (-0.55,1.5) {$\si{PD}_{1}$};
        \node[text width=3cm] at (1.7,-2.8) {$UE$};
        \node[text width=3cm,rotate=-20] at (-1.95,3.2) {\footnotesize$\si{OCIR}_{1}$};
        \node[text width=3cm,rotate=-20] at (0,3.2) {\footnotesize$\si{OCIR}_{2}$};
        \node[text width=3cm] at (0.5,4.2) {$\cdots$};
        \node[text width=3cm] at (3.7,2) {$\cdots$};
        \node[text width=3cm,rotate=-20] at (2.8,3.2) {\footnotesize$\si{OCIR}_{Q}$};
    \end{tikzpicture}
  \caption{ OCIR-based localization system model studied in this work for a room containing  $Q$ infrared PDs located on the ceiling. The UE is equipped with an infrared LED transmitter located at $(x,y,z)$, where $z$ stands for the height.}
  \label{Fig:Big_Pic}
\end{figure}

This section describes the VLC system model considered for OCIR-based localization, including the system setup, optical wireless channel, and optical signaling. 
The OCIR-based localization system configuration used in this work is depicted in Fig. \ref{Fig:Big_Pic}. 

In real-world scenarios, the localization system is part of a VLC access-point in which the infrared photodetectors (PDs) capture uplink signals from user equipment (UE) devices. We consider $Q$ uplink PDs facing vertically downwards and located on the ceiling at positions $(x^{(q)}, y^{(q)}, z^{(q)})$, where $q \in\left \{1, \cdots, Q\right \}$ and $z^{(q)}$ indicates the ceiling height. The proposed centralized localization algorithm collects all PD observations in the VLC access-point. The UE is equipped with an infrared LED transmitter facing vertically upwards, located at a fixed height and at horizontal position $\boldsymbol{\theta} = [x,y]$ with respect to the center of the room as the origin.

We adopt the system model described in \cite[Sec.~3.2]{Performance_limit_hosseinianfar} and thus forgo a detailed mathematical description of well-understood concepts like optical pulse train transmission and optical wireless channel model.  Fingerprinting-based positioning was also described in detail in \cite[Sec.~3.2]{Performance_limit_hosseinianfar}, so the concept is not repeated here. We instead focus our presentation on an extended version of the received signal used as the input to the ANN.  Table \ref{TableI} presents the definition of parameters used in our description of both the VLC system and ANN.

\begin{table}[ht]
\centering
\caption{System parameters' definition}
\label{TableI}
\setlength\extrarowheight{-3pt}
\begin{tabular}{ll}
\hline \hline \textbf{VLC System Parameters} & \textbf{Definition} \\ \hline
$\mathcal{E}_{\text{p}}$ & optical pulse energy\\ 
$T_{\text{p}}$ & pulse repetition time interval\\
$R_{\text{p}} = 1/T_{\text{p}}$ & pulse repetition rate\\
$G(t)$ & Gaussian-shaped pulse \\
$\boldsymbol{\theta}_i = [x_i,y_i]$ & $i$th UE location in fingerprinting dataset\\
$N_{\text{s}}$ & number of samples per receiver
\\
$f_s$ & sampling rate \\
$N_{\text{p}}$ & number of transmitted pulses
\\
$Q$ & number of uplink PDs\\
$\boldsymbol{r}^{(q)}(\boldsymbol{\theta}_i)= [r_{1}^{(q)},\cdots,\boldsymbol{r}_{N_{\text{s}}}^{(q)}]$ & $q$th receiver sample vector\\
 \hline
\textbf{Neural Net. Parameters} & \textbf{Definition} \rule{0pt}{8pt}\\ \hline
$\mathbf{R}(\boldsymbol{\theta}_i)= [\boldsymbol{r}^{(1)}(\boldsymbol{\theta}_i),\cdots,\boldsymbol{r}^{(Q)}(\boldsymbol{\theta}_i)]$ & $i$th ANN input vector \\
$(\mathbf{R}(\boldsymbol{\theta}_i) , \boldsymbol{\theta}_i)$ & $i$th input-label pair\\
$M$ & mini-batch size \\
$B$ & number of mini-batches per epoch \\
$n^{[\ell]}$ & number of neurons in the $\ell$th layer \\
$\mathbf{W^{[\ell]}} : (n^{[\ell]},n^{[\ell-1]})$ & coefficient vector of the $\ell$th network layer\\
$\mathbf{b^{[\ell]}} : (n^{[\ell]},1)$ & bias vector of the $\ell$th network layer \\
\hline \hline
\end{tabular}
\end{table}

As part of our optical signaling assumptions, we consider narrow time-width pulses transmitted by the UE, which enable the system to capture a high temporal resolution response profile for the optical wireless channel. Owing to practical constraints such as LED/laser diode peak power and eye-safety, the UE sends a train of optical pulses. The required optical energy for an accurate localization at the receiver is therefore split into a number of transmitted pulses, captured gradually during transmission. A detailed mathematical description of the transmitted pulse train is given in \cite[Eq.~(1)]{Performance_limit_hosseinianfar}. Optical orthogonal codes have to be transmitted by users in a multiuser case, making them distinguishable at the receiver \cite{salehi1989}.

\subsection{Optical Wireless Channel}
To model the optical wireless channel between the UE and the $q$th PD, we deploy the ray-tracing model in \cite[Eq. (1)]{lee2011indoor} expressed as
\begin{equation}
h_{\text{ch}}^{(q)} (\boldsymbol\theta,t)=h_{\text{LOS}}^{(q)} (\boldsymbol\theta,t)+ h_{\text{NLOS}}^{(q)} (\boldsymbol\theta,t),
\label{EQ:Chennel}
\end{equation}
where $h_{\text{LOS}}^{(q)}(\boldsymbol\theta,t)$ is the LOS component, and $h_{\text{NLOS}}^{(q)} (\boldsymbol\theta,t)$ is the NLOS component captured by the $q$th receiver. As shown in \cite{barry1993simulation}, one-bounce reflections play the most impotent role in defining the OCIR. Bounces beyond the first-bounce have less power and propagate in time; therefore, their ability to impair OCIR-based localization is negligible. An investigation of the degradation in performance due to multiple bounces is left for future studies. In Fig.~\ref{Fig:Big_Pic}, the OCIRs $h_{\text{ch}}^{(q)} (\boldsymbol\theta,t)$ captured at three different uplink PDs  locations for a UE at location $\boldsymbol\theta$ inside the room are shown in the green, red, and blue curves above the room. These OCIRs correspond to one-bounce reflections in an empty room obtained through \eqref{EQ:Chennel} with parameters given in \cite[Table I]{Performance_limit_hosseinianfar}. The multiple spikes in the example OCIRs can be associated with the reflections from each wall.


\subsection{Received Signal}
Using a Gaussian-shaped pulse,\footnote{Without loss of generality, a Gaussian-shaped pulse is chosen  due to its vast application in optics \cite{Miller:20}.} the optical source emits a signal that can be expressed as in \cite[Eq.~(1)]{Performance_limit_hosseinianfar}. The electrical signal is received by the $q$th receiver in the form of  \cite[Eq.~(2)]{Performance_limit_hosseinianfar}. The energy at receiver $q$ from the $N_{\text{p}}$ temporal pulses comprising the transmitted signal  are  added to  generate the signal
\begin{align}
    r^{(q)}(\boldsymbol\theta,t)&= \rho N_{\text{p}} R_{\text{p}}\mathcal{E}_{\text{p}}\left (G(t)*h_{\text{LED}}*h_{\text{ch}}^{(q)} (\boldsymbol\theta)*h_{\text{PD}}^{(q)}  \right ) (t)
   +\sqrt{N_{\text{p}}}n^{(q)}(t),
    \label{EQ:AV_rec_out}
\end{align}
in which $\rho$ is the PD responsivity, the energy per pulse is $\mathcal{E}_{\text{p}}$, $R_{\text{p}} = 1/T_{\text{p}}$ is the pulse repetition rate, where $T_{\text{p}}$ is the pulse repetition period, $G(t)$ the Gaussian-shaped pulse defined in \cite{Performance_limit_hosseinianfar}, $h_{\text{PD}}^{(q)}$ the low-pass filter that models the $q$th PD, and $h_{\text{LED}}$ is the LED impulse response. $(*)$ denotes the convolution operation. $T_{\text{p}}$ needs to be longer than the maximum length of the OCIR propagation, which is within the range of $50~\si{ns}$, i.e., a total one-bounce reflection path length of $15~\si{m}$ for typical room dimensions considered in this paper. In \eqref{EQ:AV_rec_out}, $n^{(q)}(t)$ is additive zero-mean Gaussian noise, where the factor  $\sqrt{N_{\text{p}}}$ accounts for the $N_{\text{p}}$ independent and identically distributed (i.i.d) zero-mean terms in the pulse-train. In an optical wireless environment, ambient light is the predominant noise source, adding shot noise with variance calculated based on \cite[Eq. 4]{Performance_limit_hosseinianfar}. For more details, the reader is referred to the block diagram of the positioning system depicted in \cite[Fig.~2]{Performance_limit_hosseinianfar}.
%
%

We consider a discrete sampled version of the received signal in \eqref{EQ:AV_rec_out} with a sampling rate of $f_s$, represented as $\boldsymbol{r}^{(q)}(\boldsymbol{\theta}_i)= [r_{1}^{(q)},\cdots,\boldsymbol{r}_{N_{\text{s}}}^{(q)}]$ for PD $q$,  as the input to the ANN. $N_{\text{s}}$ is the resulting number of samples within the observation window of  $50~\si{ns}$, the maximum propagation time for a typical room considered in this paper. The ultimate fingerprinting vector is a supervector resulting from concatenating these discrete vectors, i.e., $\mathbf{R}(\boldsymbol{\theta}_i) = [\boldsymbol{r}^{(1)}(\boldsymbol{\theta}_i),\cdots,\boldsymbol{r}^{(Q)}(\boldsymbol{\theta}_i)]$, where $i$ is the index of an input-label pair  $(\mathbf{R}(\boldsymbol{\theta}_i),\boldsymbol{\theta}_i)$.

Classical communication  and signal processing theorists would solve this localization problem using the well-known optimal maximum likelihood approach, which can be defined as
\begin{equation}
\underset{\boldsymbol\theta}{\text{argmax}} f_{X}( \mathbf{R}(\boldsymbol\theta)|{\boldsymbol\theta}),
\label{CRB_EQ2}
\end{equation}
where $f_{X}(\mathbf{R}(\boldsymbol\theta)|{\boldsymbol\theta})$ denotes the joint probability density function of the fingerprinting supervectors elements for a UE located at $\boldsymbol\theta$.
The optimum solution is, however, numerically prohibitive considering the nonlinear relation between the OCIR samples and the UE location. Our goal in this paper is to evaluate the effectiveness of using the OCIR for localization. Therefore, we resort to an ANN as a powerful algorithm for regression and nonlinear prediction. Comparing the ANN to the optimal approach and suboptimal heuristics (other than our benchmark, trilateration) is relegated to future work.

\section{Artificial Neural Network}\label{DL.Sec}

\begin{figure}[ht]
	\centering
    \begin{tikzpicture}
		\node(image) at (0,0) {\includegraphics[width=3.4in]{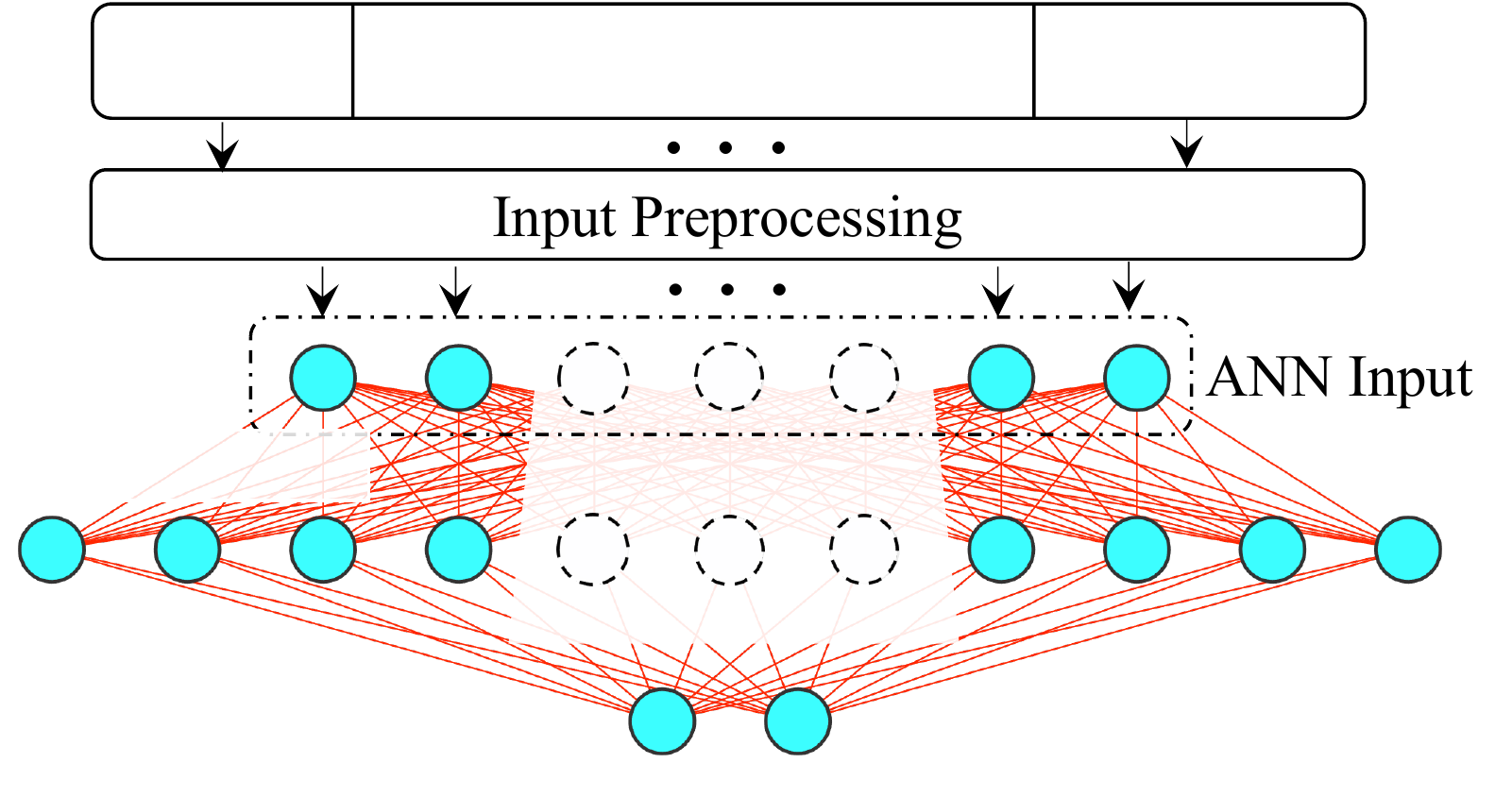}};
        \node[text width=3cm] at (-1.95,1.9) {$\boldsymbol{r}^{(1)}(\boldsymbol{\theta}_i)$};
        \node[text width=4cm] at (-0.2,1.9) {$\boldsymbol{r}^{(q)}(\boldsymbol{\theta}_i)= [r_{1}^{(q)},\cdots,\boldsymbol{r}_{N_{\text{s}}}^{(q)}]$};
        \node[text width=3cm] at (3.5,1.9) {$\boldsymbol{r}^{(Q)}(\boldsymbol{\theta}_i)$}; 
        \node[text width=3cm] at (0.45,-.45) {$\leftarrow Q \times N_{\text{s}}\rightarrow $}; 
        \node[text width=3cm] at (-2.5,-.45) {$ \left \langle  \mathbf{W^{[0]}},\mathbf{{b^{[0]}}}\right \rangle$};
        \node[text width=3cm] at (-2.5,-1.3) {$ \left \langle  \mathbf{W^{[1]}},\mathbf{{b^{[1]}}}\right \rangle$};
        \node[text width=3cm] at (0.7,-1.3) {$\leftarrow n^{[1]} \rightarrow $}; 
        \node[text width=3cm] at (0.4,-2.25) {$\boldsymbol{\hat{\theta}} \: \left \langle  \hat x~~,~~\hat y \right \rangle$};
    \end{tikzpicture}
		\caption{ ANN Architecture and notation definition}
		\label{Fig:ANN}
\end{figure}
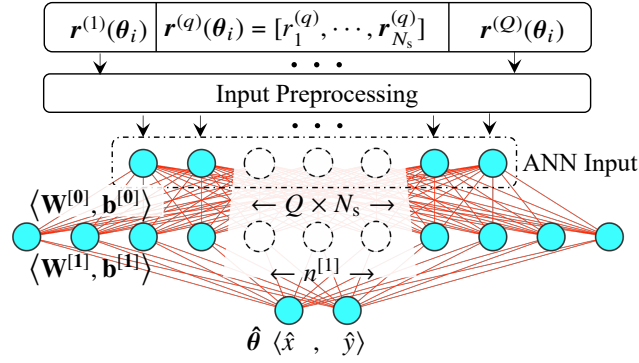

This section describes an ANN that can learn how to predict the UE's position. An artificial feedforward neural network (FNN)  with layout as illustrated in Fig.~\ref{Fig:ANN} is employed. After describing our data preprocessing, we discuss our FNNs, model training, and hyperparameter tuning, fundamental to the ANN framework. 

\subsection{Data Preprocessing}
\todo[inline,disable]{This is an important step because 1) data is noisy, 2) multipath ref. and LOS are small components to directly fed to the net.}
\todo[inline,disable]{During our tunning procedure we noticed standardization, outperform almost twice normalization. which makes sense as it makes the NLOS stronger components to be learned}
Normalization and standardization are two popular methods of preprocessing the data before inputting to the ANN.  Our numerical results showed that the ANN localization accuracy using standardization instead of normalization  improves the algorithm's performance by almost a factor of two, which makes sense because the parts of the fingerprinting supervector that contain the multipath information are emphasized. Therefore, in our numerical results we apply the common standardization technique that results in zero mean and unit variance data. We calculate the sample mean $\bar{R}$ and sample standard deviation $\mathcal{S}$ of the entire dataset over all input vector elements (all OCIR time samples  and fingerprinting supervectors)
, to create the standard vector
\begin{equation}
    \tilde{\mathbf{R}}_i(\boldsymbol{\theta}_i) = \frac{\mathbf{R}(\boldsymbol{\theta}_i)-\bar{R}}{\mathcal{S}}.
\end{equation}

\subsection{FeedForward Network}

As its name suggests, a feedforward neural network is a neural network in which the signal simply moves in one direction, from the inputs to the outputs. It is composed of one input layer, one or more hidden layers, and one output layer. Apart from the output layer, each layer is composed of neurons, fully connected to the next layer. In this paper, we consider a feedforward neural network with single hidden layer where the relation between input fingerprinting supervector and the output 2-D location can be written as

\begin{equation}\label{FFN}
\boldsymbol{\hat{\theta}}= \mathbf{W^{[1]}}\cdot\Phi(\mathbf{W^{[0]}}\cdot\tilde{\mathbf{R}}+\mathbf{b^{[0]}})+\mathbf{b^{[1]}},
\end{equation}
where $\mathbf{W^{[\ell]}}$ and $\mathbf{b^{[\ell]}}$ denote the coefficient vector and bias vector of the $\ell$th network layer, respectively, (as shown in Fig. \ref{Fig:ANN}). { $\Phi(\cdot)$ denotes a nonlinear activation function, described in detail in Section \ref{Sec:hyper_par} below.}

\subsection{Model Training}


For training the ANN, we consider an objective function based on the mean-square error (MSE) between the feed-forward network prediction (the estimated UE position inside the room) and the corresponding true UE location, plus a small L1 regularization term. The objective function is calculated as
\begin{align}\label{MSE}
    & \mathcal{L}\left(\boldsymbol{\hat{\theta}},\boldsymbol{\theta}\right )=  \frac{1}{BM}\sum_{b=1}^{B}\sum_{i=1}^{M}{\epsilon_{b,i}}, \quad \rm{where} \nonumber \\
    & \epsilon_{b,i}=\frac{1}{2}[\boldsymbol{\hat{\theta_i}}-\boldsymbol{\theta_i}][\boldsymbol{\hat{\theta_i}}-\boldsymbol{\theta_i}]^T+0.001\sum_{\ell=0}^1\left \| \mathbf{W^{[\ell]}} \right \|,  \quad \left \| \mathbf{W^{[\ell]}} \right \| = \sum_{i=1}^{n^{[\ell-1]}}\sum_{j=1}^{n^{[\ell]}}\left |  w_{i,j}^{[\ell]}\right |,
\end{align}
and $w_{i,j}^{[\ell]}$ are the elements of the $\ell$th layer coefficient matrix $W^{[\ell]}$. $M$ is the size of the mini-batch (the common subset of training data that is used for each training iteration.) $B$ denotes the number of mini-batches per epoch, i.e., the steps where the training algorithm passes the entire training data through the network. $\boldsymbol{\theta_i}$ in \eqref{MSE}  is the $i$th sample of the UE location, while $\boldsymbol{\hat{\theta_i}}$ is the $i$th sample of the UE location predicted by the ANN. 

\subsection{Activation Function and Hyperparameter Tuning}
\label{Sec:hyper_par}
\todo[inline,disable]{The ANN algorithm outperform almost twice in no-noise 1 PD scenario. The lurning curve is also less noisy when we apply l1l2 regularization.}

In this work, we  use the scaled exceptional linear unit (SELU) activation function defined as \cite{klambauer2017self} 
\begin{equation}\label{selu}
    \Phi(x) = \text{SELU}(x) = \lambda\begin{cases}
  x \quad &\text{ if } x > 0 \\ 
  \alpha e^{x}-\alpha  &\text{ if } x\leq 0.
\end{cases}
\end{equation}
{The authors of \cite{klambauer2017self} showed that SELU substantially outperforms other least rectified linear unit (ReLU) activation functions, the most often deployed functions. A self-normalized neural network, i.e., one in which overfitting is guaranteed to not occur, can be achieved using SELU.}

The design and training of ANNs is an iterative procedure that is case-dependent. While there are no theoretical criteria to optimize the design of hyperparameters, empirical metrics can help tune the performance of the system. In our case, the hyperparameter $\lambda\alpha$ determines the value that the SELU activation function approaches when $x$ in  \eqref{selu} is a large negative number. As can be seen, \eqref{selu} can take on positive and negative values, which allows the activation function to control the mean. The activation function has also the ability to dampen the variance if it is too large in the previous layer or increase it if it is too small  by defining a slope larger than one, $\lambda>1$.

\begin{figure}[ht]
\centering
    \begin{tikzpicture}
		\node(image) at (0,0) {\includegraphics[width=3.4in]{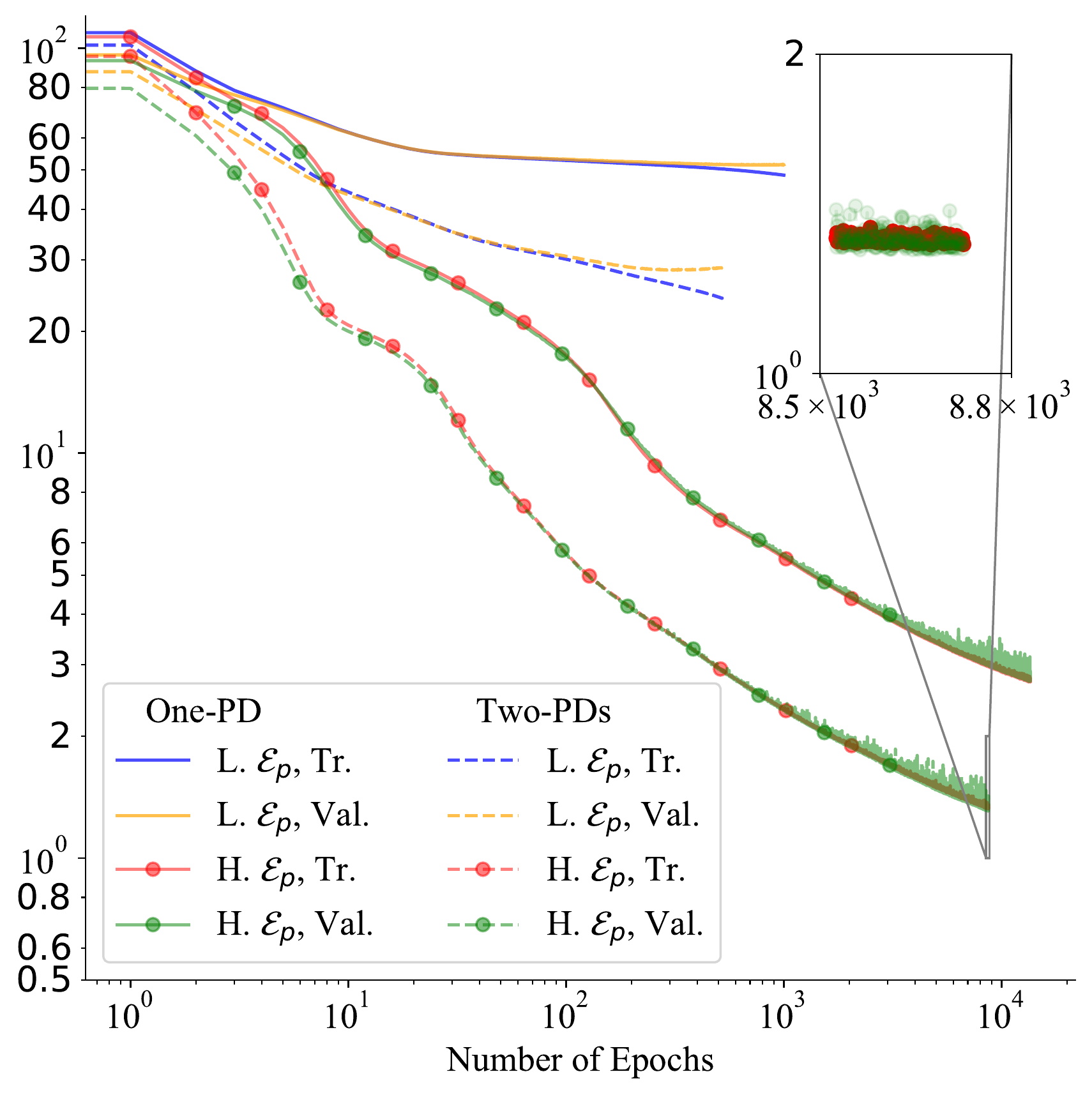}};
        \node[text width=5cm,rotate = 90] at (-4.5,1.9) {\footnotesize{$\sqrt{\mathcal{L}\left(\boldsymbol{\hat{\theta}},\boldsymbol{\theta}\right )} \approx RMSE$ (cm)}};

    \end{tikzpicture}
\caption{Learning curve for low ($\mathcal{E}_{\text{p}} = 0.01~\mu J$) and high ($\mathcal{E}_{\text{p}} = 10~\mu J$) pulse energy cases and for one and two PDs scenarios; for parameters $N_{\text{p}}=1000$, $R_{\text{p}}=100~\si{KHz}$, $f_s = 500~\si{Msps}$ ($N_{\text{s}} = 30$) and optical pulse-width of $10 \si{ns}$. The ANN trained based on the hyperparameters listed in Table \ref{TableII} } 
 \label{fig:LearningCurve}
\vspace{-.15 in}
\end{figure}

Fig.~\ref{fig:LearningCurve} illustrates the learning curves for one and two PD cases and two different pulse energies.\footnote{Based on the $\si{IEC}-62471$ standard for LED eye-safety, we calculated that $\mathcal{E}_{\text{p}}= 10~\mu J$ is within the $10~\si{cm}$ eye-safety range using a single LED at the transmitter.} Our learning algorithm exploits an early-stopping technique where the algorithm stops training when the validation root mean squared error (RMSE) does not  decrease further for 200 epochs; this technique is one of the best regularization methods to avoid overfitting \cite{ruder2016overview}. For the low-energy cases, the algorithm stops around $1000$ epochs. It is obvious from the figure that, in the low-energy cases, the training curves continue to decrease while the verification curves flatten. This is an example where the ANN is well-trained and continuing the training would lead to overfitting to noisy fingerprints. Conversely, for higher pulse-energy levels, the algorithm continues to train for almost $10000$ epochs. The algorithm input is less noisy, and therefore the ANN can extract more features from the provided OCIR vector, benefiting from longer training. 

\todo[inline,disable]{horizontal limit of 3 cm and explain why.}
\begin{table}[ht]
\centering
\caption{ANN Hyperparameter Numerical Values}
\label{TableII}
\setlength\extrarowheight{-4pt}
\begin{tabular}{ll}
\hline \hline \vspace{-10pt}\\ 
\textbf{Common Parameters} & \textbf{Value} \\ \hline \vspace{-10pt}\\  
SELU parameters & $\lambda =1.05$, $\alpha = 1.67$ \\
$M$ & $128$ \\ 
$B$ & $300$ \\
$n^{[0]}$ (input size)  & $Q\times N_{\text{s}}$ \\
$n^{[1]}$ & $400$ \\
number of training epochs  & $100000$ (early-stopping enabled) 

\\ \hline \hline
\end{tabular}
\end{table}

\section{Numerical Results and Discussion}\label{Numerical.Sec}

In this section, we present a numerical analysis of the performance of the OCIR-based localization using our ANN algorithm and compare it to conventional RSS-based linear trilateration methods. We also analyze the effectiveness of the feature extraction and localization versus different temporal resolutions, as well as the effect of the optical pulse-width. 
We deploy the VLC system as illustrated in Fig. \ref{Fig:Big_Pic} with system parameters listed in \cite[Table I]{Performance_limit_hosseinianfar}, except that the reflecting element area is modified to $A_{\text{ref}} = 1\times 1$ mm$^2$ to enhance the intrinsic temporal resolution of our simulation. The fingerprinting map for localization is generated by calculating the OCIR at different PDs for different UE locations covering the entire room on a $2$ cm grid. A temporal resolution of $0.5$ ns is considered for the OCIR function, which is far beyond the optical wireless channel bandwidth of $200~\si{MHz}$  presented in \cite{Performance_limit_hosseinianfar}. We randomly select samples of  OCIR vector/location pairs for training, verification, and testing the network using the proportions $0.64$, $0.16$, and $0.2$, respectively. Each data point presented in this section is an average of the root mean squared error (RMSE) when we train the ANN five times from scratch (called {\em resampling with five repeats}).

In the following, we first investigate the effect of the optical pulse energy on the localization accuracy. We then demonstrate the localization accuracy as a function of the receiver sampling rate and optical pulse-width.


\subsection{Accuracy versus Optical Pulse Energy $\mathcal{E}_{\text{p}}$}         
To analyze the localization accuracy as a function of the optical pulse energy, we focus on two different receiver sampling rates of $f_s = 50~\si{Msps}$ and $f_s = 500~\si{Msps}$ and a Gaussian optical pulse shape with two different pulse-widths of $1~\si{ns}$ and $10~\si{ns}$. We benchmark the proposed OCIR-based ANN algorithms against two different algorithms that rely only on DC RSS inputs from three PDs: linear trilateration with wall reflections in the simulation (employing the standard closed-form expression relating the LOS RSS to the distance \cite{lee2011indoor}) and an ANN-enabled algorithm where the unknown multipath reflection can be learned and used for localization.

\todo[inline, disable]{Redundant paragraph here, use details somewhere at the begining:This part investigates the effectiveness of deploying high temporal resolution of multipath reflection on localization performance. For this purpose, we consider the temporal resolution of $0.5$ ns which is far beyond the optical wireless channel bandwidth of $200 MHz$ as we present in \cite{Performance_limit_hosseinianfar}. We also consider high spatial resolution spatial, where the entire dataset with a spatial resolution of $2$ cm is deployed for training and testing the ANN.}

\begin{figure*}[t]
\begin{center}
\begin{tikzpicture}[x=0.2cm,y=0.2cm,scale=1,spy using outlines={rectangle,lens={scale=3}, size=3cm, connect spies}]
\begin{groupplot}[group style={
group name=my plots, group size=3 by 2,horizontal sep=0.8cm,vertical sep=35pt},
grid,
 grid style = {
    dash pattern = on 0.05mm off 1mm,
    line cap = round,
    black,
    line width = 0.2mm
  },
 ylabel style={yshift=-0.15cm},
axis y line*=left,
axis x line*=bottom,
width=6.75cm,height=7cm]
\nextgroupplot
[
ymin=0.5,ymax=100, 
xmin=0.01,xmax=10,
xmode = log,
ymode = log,
xlabel=\vspace{0.1cm} Optical Pulse Energy $\mathcal{E}_{\text{p}} (\mu J)$ , 
ylabel={RMSE (cm)},
xtick={0.01,0.05,0.1,0.5,1,5,10},
ytick={0.5,1,2,4,10,20,40,80},
xticklabels={
\hspace{0.2cm}$0.01$,
\hspace{-0.2cm}$0.05$,
$0.1$, 
$0.5$,  
$1$,
$5$,
$10$,
}, 
yticklabels={0.5,1,2,4,10,20,40,80},
]

\addplot [orange, line width=0.7mm, dotted]table[x index=0,y index=1] {Dat/Fig4/ANN_Net_400_1ns_vsOPT_loglog_July7_2022_RMSE_Tri_ref_50cm_Edge.dat};

\addplot [cyan, line width=0.5mm, solid]table[x index=0,y index=1] {Dat/Fig4/ANN_Net_400_1ns_vsOPT_loglog_July7_2022_ANN_1PD_50MSPS.dat};

\addplot [cyan, line width=0.3mm, mark=*]table[x index=0,y index=1] {Dat/Fig4/ANN_Net_400_1ns_vsOPT_loglog_Aug7_2022_ANN_1PD_500MSPS.dat};

\addplot [red, line width=0.5mm, solid]table[x index=0,y index=1] {Dat/Fig4/ANN_Net_400_1ns_vsOPT_loglog_July7_2022_ANN_2PDs_50MSPS.dat};

\addplot [red, line width=0.3mm, mark=*]table[x index=0,y index=1] {Dat/Fig4/ANN_Net_400_1ns_vsOPT_loglog_Aug7_2022_ANN_2PDs_500MSPS.dat};

\addplot [black, line width=0.5mm, dashed]table[x index=0,y index=1] {Dat/Fig4/ANN_Net_400_10ns_vsOPT_loglog_July7_2022_ANN_3PDs_Enhanced_Tri.dat};











\nextgroupplot
[legend style={at={(-0.75,0.12)},anchor=west,font=\scriptsize, row sep=-0.1cm, inner sep=0cm},
legend columns=3,
ymin=0.5,ymax=100, 
xmin=0.01,xmax=10,
xmode = log,
ymode = log,
xlabel=Optical Pulse Energy $\mathcal{E}_{\text{p}} (\mu J)$, 
xtick={0.01,0.05,0.1,0.5,1,5,10},
ytick={0.5,1,2,4,10,20,40,80},
xticklabels={
$0.01$,
\hspace{-0.2cm}$0.05$,
$0.1$, 
$0.5$,  
$1$,
$5$,
$10$, 
}, 
yticklabels={ , , , , , , , ,},
]

\addlegendimage{white, line width=0.3mm}
\addlegendimage{white, line width=0.3mm}
\addlegendimage{white, line width=0.3mm}
\addlegendimage{orange, dotted, line width=0.6mm}
\addlegendimage{cyan, line width=0.5mm}
\addlegendimage{cyan, mark=*, line width=0.5mm}
\addlegendimage{black,dashed, line width = 0.6mm}
\addlegendimage{red, line width=0.5mm}
\addlegendimage{red, mark=*, line width=0.5mm}

 \addlegendentry{\hspace{-.5cm}\textbf{Three DC RSS}$\quad\quad$}
 \addlegendentry{\hspace{-.5cm}$\mathbf{f_s = 50~\si{Msps}}\quad\quad$}
 \addlegendentry{\hspace{-.5cm}$\mathbf{f_s = 500~\si{Msps}}\quad\quad$}
\addlegendentry{Linear Tri-Lat.}
\addlegendentry{1 PD}
\addlegendentry{1 PD}
\addlegendentry{ANN-enabled}
\addlegendentry{2 PDs}
\addlegendentry{2 PDs}

\addplot [orange, line width=0.7mm, dotted]table[x index=0,y index=1] {Dat/Fig4/ANN_Net_400_10ns_vsOPT_loglog_July7_2022_RMSE_Tri_ref_50cm_Edge.dat};

\addplot [cyan, line width=0.5mm, solid]table[x index=0,y index=1] {Dat/Fig4/ANN_Net_400_10ns_vsOPT_loglog_July7_2022_ANN_1PD_50MSPS.dat};

\addplot [cyan, line width=0.3mm, mark=*]table[x index=0,y index=1] {Dat/Fig4/ANN_Net_400_10ns_vsOPT_loglog_Aug5_2022_ANN_1PD_500MSPS.dat};

\addplot [red, line width=0.5mm, solid]table[x index=0,y index=1] {Dat/Fig4/ANN_Net_400_10ns_vsOPT_loglog_July7_2022_ANN_2PDs_50MSPS.dat};

\addplot [red, line width=0.3mm, mark=*]table[x index=0,y index=1] {Dat/Fig4/ANN_Net_400_10ns_vsOPT_loglog_Aug5_2022_ANN_2PDs_500MSPS.dat};

\addplot [black, line width=0.5mm, dashed]table[x index=0,y index=1] {Dat/Fig4/ANN_Net_400_10ns_vsOPT_loglog_July7_2022_ANN_3PDs_Enhanced_Tri.dat};

\end{groupplot}

\node at (22,25) {(a)};
\node at (30+23,25) {(b)};

\end{tikzpicture}
\end{center}
\setlength{\belowcaptionskip}{-12pt}
\vspace*{-3mm}
\caption{RMS positioning error as a function of the optical transmit pulse energy for one and two PD scenarios, compared with the three PD benchmarks, for parameters $N_{\text{p}}=1000$, $R_{\text{p}}=100~\si{KHz}$, and optical pulse-width of (a)  $1~\si{ns}$ and (b) $10~\si{ns}$} 
\label{Fig:ANN_Net_100vsOPT}
\end{figure*}
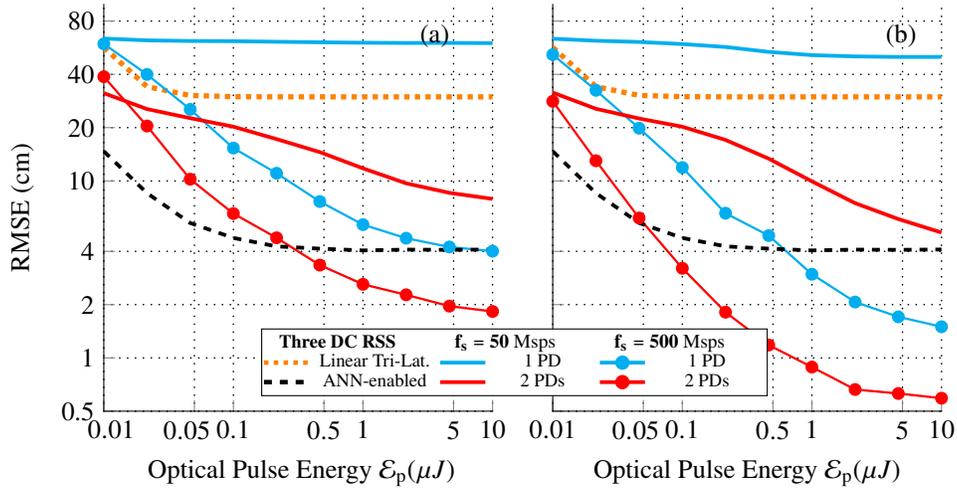

Fig. \ref{Fig:ANN_Net_100vsOPT} (a) illustrates the RMSE  of the proposed ANN localization algorithm versus different levels of optical pulse energy for an optical pulse-width of $1~\si{ns}$. As the pulse energy increases, so does the algorithm accuracy, up to a point. By using a higher temporal resolution of the received signal, we can take advantage of more detailed NLOS temporal patterns to improve the localization accuracy. One photodetector is unable to accurately locate the UE when the received sampling rate is low, but can achieve a $4$ cm accuracy with faster receiver sampling. A two-PD system can achieve a $9$ cm accuracy even with the lower receiver sampling rate.

Strong reflections coming from walls are seen as destructive noise in conventional trilateration \cite{gu2016impact}, which is why it fails to achieve an accuracy below $30~\si{cm}$.\footnote{The conventional linear trilateration results were calculated using  system parameter given in \cite[Table I]{Performance_limit_hosseinianfar}, which are close to the results provided in \cite{gu2016impact}. To compare our proposed method with the previous ones fairly, we excluded a $0.5~\si{m}$ edge from the walls where severe reflection leads to significant localization errors for conventional trilateration.} When we use the high temporal resolution OCIR as a fingerprinting map with two PDs, the ANN localization algorithm outperforms the conventional RSS techniques by about one order of magnitude. Applying the ANN as an empirical feature selection and localization technique also enables it to improve the performance of the original OCIR-based localization presented in \cite{Performance_limit_hosseinianfar}

If three PDs are available to collect information, the unknown multipath reflection can be learned using an ANN even in the simple DC RSS measurement case, enhancing the accuracy to $5~\si{cm}$, much better than simple trilateration, as shown in Fig. \ref{Fig:ANN_Net_100vsOPT} (a). 
Using one PD and $f_s = 500~\si{Msps}$, the OCIR ANN algorithm asymptotically reaches the performance of the ANN-enabled three-DC-RSS algorithm. 
In addition, the algorithm with two PDs and the high sampling rate outperforms the ANN-enabled three-DC-RSS scenario by a factor of two. 

For both one and two PDs cases, the low and high sampling rate curves cross each other. This means that in a low-energy scenario with a higher sampling rate, the ANN collects more noise and less signal energy per sample, and therefore performs worse than the lower sampling rate case.

The localization accuracy is improved for an optical pulse-width of $10~\si{ns}$ shown in Fig. \ref{Fig:ANN_Net_100vsOPT}~(b), as compared with the $1~\si{ns}$ pulse-width results shown in Fig. \ref{Fig:ANN_Net_100vsOPT}~(a)---one would expect a higher  accuracy using narrower pulse-widths. This behavior is due to a trade-off between enriching the input vector by allowing the dominant LOS received energy to span more vector dimensions, i.e., temporal samples, and losing temporal resolution due to sending wider optical pulses. In the narrow pulse case, the LOS is received in one or two temporal samples when a high receiver sampling rate is used; thus, these cannot have their warranted effect on the ANN performance.  In other words, while we lose the temporal resolution of the NLOS pattern by transmitting a wider pulse, we create a richer input vector that enables the ANN algorithm to distinguish the received signal in more effective dimensions.

\subsection{Temporal Resolution Effect on the Performance}
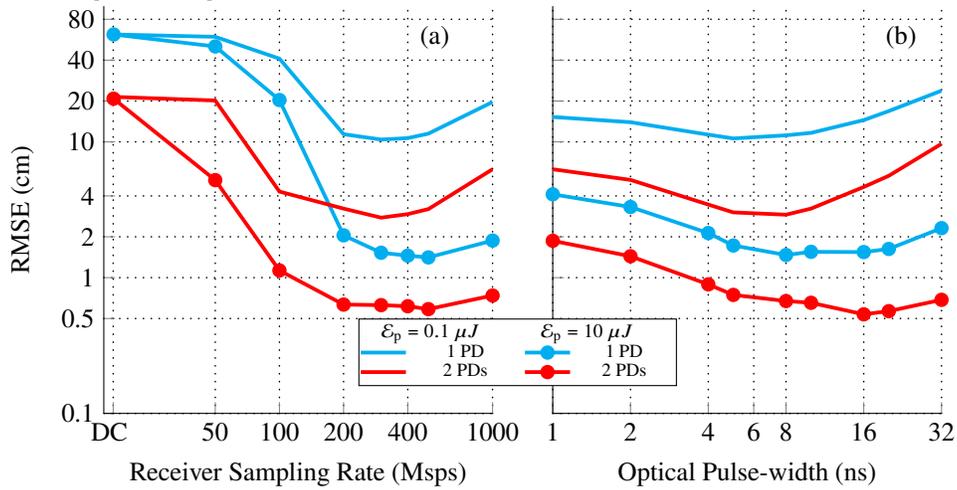
\begin{figure*}[t]
\begin{center}
\begin{tikzpicture}[x=0.2cm,y=0.2cm,scale=1,spy using outlines={rectangle,lens={scale=3}, size=3cm, connect spies}]
\begin{groupplot}[group style={
group name=my plots, group size=3 by 2,horizontal sep=0.8cm,vertical sep=35pt},
grid,
 grid style = {
    dash pattern = on 0.05mm off 1mm,
    line cap = round,
    black,
    line width = 0.2mm
  },
 ylabel style={yshift=-0.15cm},
axis y line*=left,
axis x line*=bottom,
width=6.75cm,height=7cm]
\nextgroupplot
[
ymin=0.1,ymax=100, 
xmin=15,xmax=1000,
xmode =log,
ymode = log,
xlabel=Receiver Sampling Rate (Msps), 
ylabel={RMSE (cm)},
xtick={16.6,50,100,200,400,1000},
ytick={0.1,0.5,1,2,4,10,20,40,80},
xticklabels={
\hspace{-0.1cm}DC,
$50$, 
$100$,  
$200$,
$400$,
$1000$,
},
yticklabels={0.1,0.5,1,2,4,10,20,40,80},
]

\addplot [cyan, line width=0.5mm, solid]table[x index=0,y index=1] {Dat/Fig5/ANN_Net_400vsfs_loglog_May30_2020_ANN_1PD_LE.dat};

\addplot [cyan, line width=0.5mm, solid, mark=*]table[x index=0,y index=1] {Dat/Fig5/ANN_Net_400vsfs_loglog_May30_2020_ANN_1PD_HE.dat};

\addplot [red, line width=0.5mm]table[x index=0,y index=1] {Dat/Fig5/ANN_Net_400vsfs_loglog_May30_2020_ANN_2PDs_LE.dat};

\addplot [red, line width=0.5mm, solid, mark=*]table[x index=0,y index=1] {Dat/Fig5/ANN_Net_400vsfs_loglog_May30_2020_ANN_2PDs_HE.dat};













\nextgroupplot
[legend style={at={(-0.5,0.15)},anchor=west,font=\scriptsize, row sep=-0.1cm, inner sep=0cm},
legend columns=2,
ymin=0.1,ymax=100, 
xmin=1,xmax=32,
xmode = log,
ymode = log,
xlabel=Optical Pulse-width (ns), 
xtick={1,2,4,6,8,16,32},
ytick={0.1,0.5,1,2,4,10,20,40,80},
xticklabels={
$1$,
$2$,
$4$, 
$6$,  
$8$,
$16$,
$32$,
}, 
yticklabels={ , , , , , , , , },
]

\addlegendimage{white, line width=0.5mm}
\addlegendimage{white, line width=0.5mm}
\addlegendimage{cyan, line width=0.5mm}
\addlegendimage{cyan, mark=*, line width=0.5mm}
\addlegendimage{red, line width=0.5mm}
\addlegendimage{red, mark=*, line width=0.5mm}

 \addlegendentry{\hspace{-0.5cm} $\mathcal{E}_{\text{p}} = 0.1~\mu J \quad\quad$}
 \addlegendentry{\hspace{-0.5cm}$\mathcal{E}_{\text{p}} = 10~\mu J\quad\quad$}

\addlegendentry{1 PD}
\addlegendentry{1 PD}
\addlegendentry{2 PDs}
\addlegendentry{2 PDs}

\addplot [cyan, line width=0.5mm, solid]table[x index=0,y index=1] {Dat/Fig5/ANN_Net_400vsPulsewidth_loglog_July7_2022_ANN_1PD_LE.dat};

\addplot [cyan, line width=0.5mm, solid, mark=*]table[x index=0,y index=1] {Dat/Fig5/ANN_Net_400vsPulsewidth_loglog_July7_2022_ANN_1PD_HE.dat};

\addplot [red, line width=0.5mm]table[x index=0,y index=1] {Dat/Fig5/ANN_Net_400vsPulsewidth_loglog_July7_2022_ANN_2PDs_LE.dat};

\addplot [red, line width=0.5mm, solid, mark=*]table[x index=0,y index=1] {Dat/Fig5/ANN_Net_400vsPulsewidth_loglog_July7_2022_ANN_2PDs_HE.dat};

\end{groupplot}

\node at (22,25) {(a)};
\node at (30+23,25) {(b)};

\end{tikzpicture}
\end{center}
\setlength{\belowcaptionskip}{-12pt}
\vspace*{-3mm}
\caption{RMS positioning error for one and two PD scenarios and for parameters $N_{\text{p}}=1000$, $R_{\text{p}}=100~\si{KHz}$ as a function of the (a) receiver sampling rate (optical pulse-width of $10~\si{ns}$), and (b) optical pulse-width (sampling rate of $500~\si{Msps}$)} 
\label{Fig:ANN_Net_100vsBW}
\end{figure*}

This section compares the effect the received signal's temporal resolution has on the OCIR-based ANN localization performance. Fig.~\ref{Fig:ANN_Net_100vsBW}~(a) illustrates the localization RMSE as a function of the receiver sampling rate for a pulse-width of $10~\si{ns}$ and two different pulse energies. Even at low sampling rates this algorithm can still benefit from multipath reflection for localization. The trade-off between the sampling rate and energy-to-noise per sample discussed above emerges for all curves in Fig. \ref{Fig:ANN_Net_100vsBW} (a) as a minimum RMSE yielding the best accuracy around sampling rate of $f_s = 200~\si{Msps}$.


Fig. \ref{Fig:ANN_Net_100vsBW} (b) illustrates the impact that the transmitted optical pulse-width can have on the localization performance. As discussed above, a narrower pulse-width does not necessarily give rise to better performance. There is, however, an optimum pulse-width corresponding to the trade-off between the input vector richness and temporal resolution.

\todo[inline,caption={},disable]{Conclusion Paragraph:
\begin{itemize}
\item The performance of the localization algorithm can outperform more than one order of magnitude compared with conventional RSS techniques when we deploy the high temporal resolution of OCIR as a fingerprinting map and only two PDs.
\item In addition, the algorithm outperforms the original OCIR-based localization presented in \cite{Performance_limit_hosseinianfar} when we use ANN as best empirical feature selection and localization technique compared with limited feature selection and sub-optimal one nearest neighbor algorithm.
\item The performance of the OCIR-based method degrades as we reduce the receiver sampling rate. However, the algorithm still outperforms the conventional RSS even when we only capture one sample per PD, i.e., by learning unknown but deterministic DC numbers represented the multipath reflection.
\end{itemize}
}
\section{Conclusion}\label{Conclusion.Sec}

The effectiveness of using LOS and multipath reflections in an optical wireless channel to construct and use a fingerprinting map for indoor localization was studied in this paper. A neural network architecture is used to extract features of the OCIR and then map them to the user equipment location.  The proposed scheme achieves a two orders-of-magnitude localization accuracy improvement compared to conventional RSS methods. The ANN technique explored in this work can exploit multipath reflections in many scenarios, including the use of one or more PDs, a range of transmitter pulse-widths, and various received sampling rates as low as collecting a single DC value per photodetector. {The OCIR-based ANN localization with two PDs outperforms the ANN-enabled RSS-based algorithm with three LOS, i.e., the algorithm is able to exploit the multipath reflection enough to compensate for losing one LOS signal due to blockage.}
The results indicate that, for the OCIR-ANN, an RMS positioning error of $5$~mm can be achieved using two PDs with an optical pulse energy of $10~\mu J$, optical pulse width of $10~\si{ns}$, and a receiver sampling rate of $500~\si{Msps}$. The OCIR-based ANN localization algorithm outperforms conventional trilateration even for LED and PD bandwidths lower than $100~\si{MHz}$.

{From a theoretical perspective, finding fundamental limits for OCIR-based methods is an interesting open problem to investigate. Though this work uses ANN as a tool for performance analysis of OCIR-based localization, investigating effective  neural networks architectures, including multiple hidden layers and different numbers of neurons, is a possible extension of this paper. Addressing practical aspects such as training with a limited dataset, simultaneous mapping and localization, and developing an experimental setup is also relegated to future work.} 


\section*{Acknowledgement}
The authors acknowledge Research Computing at The University of Virginia for providing computational resources and technical support that have contributed to the results reported within this publication. URL: https://rc.virginia.edu
\bibliography{VLC}
\end{document}